\documentclass[12pt,thmsa]{article}
\usepackage{sw20aip}

\input{tcilatex}
\input tcilatex
\begin{document}

\title{What is entanglement?}
\author{Emilio Santos \\
Departamento de F\'{i}sica. Universidad de Cantabria. Santander. Spain}
\maketitle

\begin{abstract}
I conjecture that only those states of light whose Wigner function is
positive are real states, and give arguments suggesting that this is not a
serious restriction. Hence it follows that the Wigner formalism in quantum
optics is capable of interpretation as a classical wave field with the
addition of a zeropoint contribution. Thus entanglement between pairs of
photons with a common origin occurs because the two light signals have
amplitudes and phases, both below and above the zeropoint intensity level,
which are correlated with each other.
\end{abstract}

\section{What states of light are real?}

I shall begin with the following

\begin{conjecture}
Real states of light have a positive Wigner function.
\end{conjecture}

Probably the reader does not believe that this conjecture may be true.
Therefore I shall rewrite it in a more acceptable form:

\begin{conjecture}
In all experiments where the conventional quantum interpretation involves
states of light with negative Wigner functions, the interpretation would
require only positive Wigner functions if all sources of noise, and other
nonidealities, are taken into account.
\end{conjecture}

\subsection{The quantum states of light and their Wigner representation}

The ''states'' which form the basis for the quantum optical description of
the light field are the \textit{Fock states}, which are generated by
applying the \textit{creation operators} $\hat{a}_{\mathbf{k},\lambda
}^{\dagger }$ to the hilbert-space vector 
\begin{equation}
|0\rangle =\prod_{\mathbf{k},\lambda }|0_{\mathbf{k},\lambda }\rangle \;,
\label{vac}
\end{equation}
which represents the vacuum. The whole Fock space is then spanned by the set
of vectors 
\begin{equation}
|\{n_{\mathbf{k},\lambda }\}\rangle =\prod_{\mathbf{k},\lambda }|n_{\mathbf{k%
},\lambda }\rangle =\prod_{\mathbf{k},\lambda }\frac{1}{\sqrt{n_{\mathbf{k}%
,\lambda }!}}(\hat{a}_{\mathbf{k},\lambda }^{\dagger })^{n_{\mathbf{k}%
,\lambda }}|0_{\mathbf{k},\lambda }\rangle \;,  \label{fock}
\end{equation}
which represents a state having $n_{\mathbf{k},\lambda }$ photons of wave
number \textbf{k} and polarization $\lambda $. The latter index takes either
the value 1 or 2. The most general \textit{pure state} is a superposition of
these, that is 
\begin{equation}
|\Phi \rangle =\sum \phi (\{n_{\mathbf{k},\lambda }\})|\{n_{\mathbf{k}%
,\lambda }\}\rangle \;,\;\sum |\phi |^{2}=1\;.  \label{pure}
\end{equation}
The full set of quantum states is obtained by extending the set $|\Phi
\rangle $ to \textit{mixtures} of the form 
\[
\widehat{\rho }=\sum |\Phi \rangle P_{\Phi }\langle \Phi |\;,\;0\leq P_{\Phi
}\leq 1\;,\;\sum P_{\Phi }=1\;.
\]

The Wigner function of this state is defined as 
\begin{equation}
W_{\rho }(\{\alpha _{\mathbf{k},\lambda }\})=Tr\left[ \widehat{\rho }\hat{W}%
(\{\alpha _{\mathbf{k},\lambda }\})\right] \;,  \label{wro}
\end{equation}
where $\{\alpha _{\mathbf{k},\lambda }\}$ are a set of complex variables,
representing the amplitudes of the radiation modes, and 
\[
\hat{W}(\{\alpha _{\mathbf{k},\lambda }\})=\prod_{\mathbf{k},\lambda }\frac{1%
}{\pi ^{2}}\int \exp \left[ \xi _{\mathbf{k},\lambda }(\hat{a}_{\mathbf{k}%
,\lambda }^{\dagger }-\alpha _{\mathbf{k},\lambda }^{*})-\xi _{\mathbf{k}%
,\lambda }^{*}(\hat{a}_{\mathbf{k},\lambda }-\alpha _{\mathbf{k},\lambda
})\right] d^{2}\xi _{\mathbf{k},\lambda }\;.
\]
(The integration should be performed with respect to the real and imaginary
parts of every complex variable.) For instance, the Wigner function of the
vacuum is 
\begin{equation}
\mathcal{W}_{0}(\{\alpha _{\mathbf{k},\lambda }\})=\prod_{\mathbf{k},\lambda
}(2/\pi )\exp \left( -2|\alpha _{\mathbf{k},\lambda }|^{2}\right) .
\label{wvac}
\end{equation}

In the hilbert-space representation the electric and magnetic fields of the
radiation are usually expanded in normal modes, the coefficients of the
expansion being the creation, $\hat{a}_{\mathbf{k},\lambda }^{\dagger },$
and annihilation, $\hat{a}_{\mathbf{k},\lambda },$ operators of photons. In
the Wigner representation these operators become the amplitudes of the
modes, which are random variables with a distribution given by the Wigner
function (assumed positive). For instance the expansion of the electric
field in free space is 
\begin{equation}
\mathbf{E(r},t)=\sum_{\mathbf{k},\lambda }\sqrt{\frac{2\hbar \omega }{L^{3}}}%
\func{Re}\left[ \alpha _{\mathbf{k},\lambda }\mathbf{e}_{\mathbf{k},\lambda
}\exp \left( i\mathbf{k.r}-i\omega t\right) \right] ,\omega \equiv c|\mathbf{%
k}|,  \label{field}
\end{equation}
where $\hbar $ is Planck's constant, $\mathbf{e}_{\mathbf{k},\lambda }$ the
polarization vector and $L^{3}$ the normalization volume.

The Wigner function in nonrelativistic quantum mechanics \cite{w} plays the
role of a pseudoprobability distribution; its marginals with respect to
position and momentum separately give the quantum probabilities for each of
these variables, but the function itself is not positive definite. There are
great difficulties in interpreting the Wigner function as a true probability
distribution in quantum mechanics. Nevertheless I propose that for the light
field the Wigner function may be assumed positive for all physically
realizable states. This assumption will not put too big difficulties for the
interpretation of actual experiments, as is argued in the next subsection.

\subsection{When is the Wigner function positive ?}

The Fock states $\left( \ref{fock}\right) $ form a basis appropriate for the
solution of the Maxwell equations fulfilled by the electromagnetic field.
They play the same role as, for instance, the functions $\left\{
sin(nx)\right\} $ in the solution of the diffusion equation in one
dimension, 
\[
\frac{\partial f}{\partial t}=\frac{\partial ^{2}f}{\partial x^{2}},
\]
$f(x,t)$ being the density of the diffusing matter. With boundary conditions 
$f(0,t)=f(\pi ,t)$ $=0,$ the solution is 
\begin{equation}
f(x,t)=\sum c_{n}sin(nx)\exp (-n^{2}t),  \label{series}
\end{equation}
where the coefficients c$_{n}$ may be obtained from the initial condition $%
f(x,0)$. We need the whole set of functions $\left\{ sin(nx)\right\} $ with
integer $n$ in order to be able to express the solution of the problem as a
Fourier series (this requirement is called completeness of the basis), but
it is obvious that not all series of the form $\left( \ref{series}\right) $
may represent real physical states. In particular the functions $\left\{
sin(nx)\right\} $ themselves are not positive definite (except for $n\ =1$)
and cannot represent densities. Similarly we may assume that the ''Fock
states'' do not correspond to physically realizable states (except $\left( 
\ref{vac}\right) )$ although all of them are necessary in order to represent
every possible physical state of the radiation field.

\smallskip Our assumption contrasts with the frequent interpretation of
experiments in terms of one-photon states. The one-photon state $|1\rangle $
has the Wigner function 
\[
W_{1}(\alpha )=(4|\alpha |^{2}-1)W_{0}(\alpha )\;,
\]
where $W_{0}(\alpha )$ is the vacuum Wigner function for a single mode $%
\left( \ref{wvac}\right) .$ Taking into account the complete set of modes,
including the ``unoccupied'' ones, the Wigner function may be written as 
\[
\mathcal{W}_{1}(\{\alpha _{\mathbf{k},\lambda }\})=(4|\alpha _{\mathbf{k}%
,\lambda }|^{2}-1)\;\mathcal{W}_{0}(\{\alpha _{\mathbf{k},\lambda }\}).
\]
Of course this is not always positive.

Actually the experimental situations \cite{clau1,grangier2} in which
something like a one-photon state has been reported, must necessarily
involve the observations of wave packets rather than single-mode signals.
Indeed, the latter, which fill the whole of space and time, are not at all
physical objects. A wave packet has the hilbert-space representation 
\begin{equation}
|\zeta (\mathbf{x})\rangle =\sum_{\mathbf{k},\lambda }\zeta _{\mathbf{k}%
,\lambda }e^{i\mathbf{k.x}}a_{\mathbf{k},\lambda }^{\dagger }|0\rangle \;,
\label{onephoton}
\end{equation}
where $\{\zeta _{\mathbf{k},\lambda }\}$ are a set of random, but not
independent variables satisfying 
\[
\sum_{\mathbf{k},\lambda }|\zeta _{\mathbf{k},\lambda }|^{2}=1\;,
\]
and which, furthermore, are nonzero only for a set of vectors \textbf{k}
falling within a small ellipsoidal region centred at $\mathbf{k^{\prime }}$.

The Wigner function of $\left( \ref{onephoton}\right) $ is not positive.
However, it is necessary to bear in mind that there is no way of controlling
the moment at which such a packet is emitted, in the atomic-cascade
situation used for such experiments, and this may be taken into account by
forming an appropriate mixture of such wave-packet states. We have been able
to show that the one-photon state becomes a mixture having a positive Wigner
function \cite{balt,fpl}. The proof will not be reproduced here, but a hint
may be got by realizing that the mixed state with density operator 
\[
\widehat{\rho }=\frac{1}{2}|1\rangle \langle 1|+\frac{1}{2}|0\rangle \langle
0|
\]
has a positive Wigner function, namely 
\[
\mathcal{W}_{1}(\{\alpha _{\mathbf{k}^{\prime },\lambda ^{\prime
}}\})=2|\alpha _{\mathbf{k},\lambda }|^{2}\mathcal{W}_{0}(\{\alpha _{\mathbf{%
k}^{\prime },\lambda ^{\prime }}\})\;.
\]

Incidentally, with our approach it is possible to explain, in terms of pure
waves, one of the most tramatic instances of ''corpuscular behaviour'' of
light, the anticorrelation after a beam splitter \cite{grangier2}-\cite{fpl}%
. But the wave-particle duality of light will not be treated further in this
article.

In summary I challenge the current, although rarely explicit, assumption
that all density operators represent physically realizable states. In
contrast I propose that only those states having a positive Wigner function
may be realizable. On the other hand one may ask whether there are real
states such that their (positive) probability distribution, $W$, is not a
Wigner function, that is no density operator $\widehat{\rho }$ exists
leading to $W$ via eq.$\left( \ref{wro}\right) .$ Such states might violate
the Heisenberg uncertainty inequalities which, in our approach, derive from
the existence of a minimal noise which may be controlled in experiments.
Thus I think that the answer is in the negative, in the sense that the
amount of noise in the preparation procedure cannot be reduced below the
Heisenberg inequalities limit. Indeed, it will be usually well above that
limit. In any case the question will not be studied here further \cite{fp11}.

\subsection{Quantum pure states which are real}

Pure states, of the form $\left( \ref{pure}\right) ,$ rarely have a positive
Wigner function. Indeed it is known that if the Wigner function is positive,
it is gaussian. The proof was given in \cite{hudson} for a single mode and
generalized to many modes in \cite{claverie}. In contrast, no general rule
is known in the case of quantum mixed states.

In the rest of this article we shall treat only the restricted, but
important, kind of real states having a gaussian Wigner function. For
simplicity we shall consider the ideal (i. e. unphysical) situation where
only a single mode of the field contains photons, so that the set $\{n_{%
\mathbf{k},\lambda }\}$ contains a single member, and the number states are
designated simply $|0\rangle ,|1\rangle ,|2\rangle \ldots $. The full Wigner
function will be given by the product of the single-mode function with $%
W_{0}(\alpha _{\mathbf{k},\lambda })$ for each of the ``unoccupied'' modes.
The generalization to many modes should not be difficult.

The most general single-mode gaussian Wigner function may be written in
terms of two real parameters, A and B, and a complex one, a, as follows 
\begin{equation}
W(\alpha )=\frac{\sqrt{AB}}{\pi }\exp \left\{ -A\left( \func{Re}\alpha -%
\func{Re}a\right) ^{2}-B\left( \func{Im}\alpha -\func{Im}a\right)
^{2}\right\} .  \label{gauss}
\end{equation}
If $AB=4$, $W$($\alpha )$ corresponds to a quantum pure state and if $AB>4$
to a mixed state. Values such that $AB<4$ do not provide a Wigner function,
that is no density operator exists whence $W$($\alpha )$ may obtained via $%
\left( \ref{wro}\right) $.

Among pure states the case $A=B$, that is 
\begin{equation}
W_{a}(\alpha )=(2/\pi )\mathrm{\exp }\left( -2|\alpha -a|^{2}\right) ,
\label{coh}
\end{equation}
is called \textit{coherent state}, which is an idealized form of the
continuous-wave laser if $a=0$ (and represents the vacuum $\left( \ref{wvac}%
\right) $ if $a\neq 0)$. The case $A$ $\neq B$, that is 
\[
W_{a,s}(\alpha )=(2/\pi )\exp [-2e^{2s}(\func{Re}\alpha -\func{Re}%
a)^{2}-2e^{-2s}(\func{Im}\alpha -\func{Im}a)^{2}],
\]
with $s$ real is called \textit{squeezed state}. The hilbert-space
representations of these two states are, respectively, 
\begin{equation}
|a\rangle =\mathrm{\exp }\left( a\hat{a}^{\dagger }-a{}^{*}\hat{a}\right)
|0\rangle =\sum_{n}\frac{a{}^{n}}{\sqrt{n!}}\,\mathrm{\exp }\left(
-|a{}^{2}|/2\right) |n\rangle ,  \label{coher}
\end{equation}
and 
\begin{equation}
|a,s\rangle =\exp \left( s(\hat{a}^{\dagger 2}-\hat{a}^{2})\right) |a\rangle
.  \label{squeez}
\end{equation}

States $\left( \ref{coher}\right) $ and $\left( \ref{squeez}\right) $ are
the only single-mode quantum pure states which are real according to our
criterion.

\subsection{Stochastic optics}

If all real states of light have a positive Wigner function, we may
interpret that function as an actual probability distribution of the
amplitudes of the radiation modes. Thus quantum optics becomes a disguised
stochastic theory, where the states of light are probability distributions
defined on the set of possible realizations of the electromagnetic field. We
propose the name \textit{stochastic optics} \cite{marsant} for the
stochastic interpretation of quantum optics derived from the Wigner
function. From another point of view, the stochastic interpretation provides
an explicit hidden variables theory where \emph{the amplitudes of the
electromagnetic field are the ''hidden'' variables!.}

The most dramatic consequence of stochastic optics is that the vacuum is no
longer empty, but filled with a random electromagnetic radiation having an
energy $\frac{1}{2}\hbar \omega $ per radiation mode, on the average, as is
shown in eq.$\left( \ref{wvac}\right) $. That radiation corresponds
precisely to the additional term introduced by Max Planck in his second
radiation law (see e.g. \cite{milonni} ). The picture that emerges is that
space contains a random background of electromagnetic waves providing what
we shall call a zeropoint field (ZPF).

The question whether that theory should be named \emph{classical} is a
matter of taste. It might be said classical in the sense that it is a pure
wave theory, essentially the same developed during the XIX Century. Photons
are just wavepackets, usually localized in the form of needles of radiation
(see eq.$\left( \ref{onephoton}\right) $), superimposed to the ZPF.
Nevertheless the theory departs from classical optics in the assumption that
there exists a fundamental noise, the ZPF, which cannot be eliminated even
at zero Kelvin. I prefer to remain closer to the current nomenclature and
say that \emph{stochastic optics is not a classical theory}.

\section{What states of light are classical?}

\subsection{Zeropoint field and detection theory}

The ZPF is usually not directly observable, although indirectly it may
produce observable effects, as explained below. Observable signals consist
of additional radiation on top of the ''sea'' of ZPF. Crucial for the
stochastic interpretation is the assumption that the ZPF has precisely the
same nature as signals. Therefore explaining why the ZPF is not directly
observable, e. g. by firing photon detectors, is a non-trivial problem which
will not be studied here (see, e. g., \cite{pdc7} ). In the following we
state the problem more precisely.

The hilbert-space formalism of detection is based on normal ordering, that
is putting creation operators to the left and annihilation operators to the
right. Let us begin with the ideal case where the radiation field is
represented by a single mode with amplitude $\alpha $. Then the detection
probability per unit time is 
\begin{eqnarray}
P^{q} &\propto &Tr\left[ \widehat{\rho }\hat{a}^{\dagger }\hat{a}\right] =%
\frac{1}{2}Tr\left[ \widehat{\rho }\left\{ \hat{a}^{\dagger }\hat{a}+\hat{a}%
\hat{a}^{\dagger }-1\right\} \right]  \nonumber \\
&=&\int W(\{\alpha )\left[ |\alpha |^{2}-\textstyle\frac{1}{2}\right]
d^{2}\alpha \equiv \langle |\alpha |^{2}-\frac{1}{2}\rangle _{_{W}}=n
\label{pw0}
\end{eqnarray}
where $\widehat{\rho }$ is the state of the radiation, $W(\alpha )$ the
corresponding Wigner function, and $n$ is the ''mean number of photons'' in
the state. The first equality derives from the use of the commutation
relations and the second is the passage to the Wigner representation. The
symbol $\langle \rangle _{_{W}}$ means the average of the quantity inside
weighted with the Wigner function of the state.

In the general case of many modes and working in the Heisenberg picture, it
is straightforward to get, for a point-like detector \cite{pdc1}, 
\begin{equation}
P^{q}(\mathbf{r},t)\propto \int W(\{\alpha _{\mathbf{k}}\})\left[ \mathcal{I}%
\left( \mathbf{r},t;\{\alpha _{k}\}\right) -I_{0}\right] d^{2N}\alpha _{%
\mathbf{k}}=\langle \mathcal{I}-I_{0}\rangle \;,  \label{pw1}
\end{equation}
where 
\begin{equation}
\mathcal{I}(\mathbf{r},t;\{\alpha _{\mathbf{k}}\})=c\epsilon _{0}\left| 
\mathbf{E}(\mathbf{r},t;\{\alpha _{\mathbf{k}}\})\right| ^{2}  \label{intens}
\end{equation}
is the intensity for a realization of the field at the position and time $(%
\mathbf{r},t)$. $W(\{\alpha _{\mathbf{k}}\})$ is the Wigner function of the
initial state, $N$ is the number of modes (we should take the limit $%
N\rightarrow \infty $ at some appropriate moment) and $I_{0}$ is the mean
intensity of the ZPF. (We use the ''caligraphic'' $\mathcal{I}$ for the
(random variable) intensity in order to distinguish it from the (nonrandom)
average $I_{0}.)$ Writing the electric field, \textbf{E}, in terms of the
initial amplitudes of the normal modes, $\{\alpha _{\mathbf{k}}\}$, is
usually straightforward although lengthy. It should be made for every
particular experiment (almost all performed experiments with light produced
by parametric down conversion have been studied with the Wigner function
formalism in the series of articles \cite{pdc1}-\cite{pdc4}).

Eq.$\left( \ref{pw1}\right) $ may be interpreted as stating that the
detector has a threshold so that it only detects the part of the field which
is above the average ZPF, that is the detector removes the ZPF. The quantum
rule $\left( \ref{pw1}\right) $ is just to subtract the mean, a formal
procedure which cannot be physical because it gives rise to ``negative
probabilities''. The problem is not the huge value of the zeropoint energy
(the ZPF intensity is about 10$^{5}$w/cm$^{2}$ in the visible range),
because the threshold intensity $I_{0}$ cancels precisely that intensity.
The problem lies in the fluctuations of the intensity. For the weak light
signals of typical quantum-optical experiments the fluctuations of $\mathcal{%
I}$ may be such that $\mathcal{I}<I_{0}$. This problem may be solved as
discussed elsewhere \cite{pdc7}, but it will not be considered here further.

\subsection{Classical and nonclassical states of light}

The fact that nonclassicality derives from the existence of the ZPF provides
a criterion to classify the states of the radiation field. We shall call
classical (nonclassical) those states where the ZPF is irrelevant
(relevant). More specifically we will consider a state as classical if the
total field $\mathbf{E(r,}t)$ can be decomposed into two independent parts, $%
\mathbf{E_{0}(r,}t)$ and $\mathbf{E_{1}(r,}t)$ representing ZPF and signal
respectively \cite{balt}. When this is the case, all optical phenomena are
associated with the signal alone, and the ZPF may be ignored altogether, as
is the situation in classical optics. In particular detectors remove
precisely the ZPF, see $\left( \ref{pw1}\right) $. The independence of the
fields implies that the corresponding amplitudes, $\{\alpha _{\mathbf{k}%
,\lambda }^{0}\}$ and $\{\alpha _{\mathbf{k},\lambda }^{1}\},$ are
independent random variables. If the probability densities of these are $%
W_{0}(\{\alpha _{\mathbf{k},\lambda }^{0}\})$ and $W_{1}(\{\alpha _{\mathbf{k%
},\lambda }^{1}\})$, then the density of the total field will be their
convolution, that is 
\[
W(\{\alpha _{\mathbf{k},\lambda }\})=\int W_{1}(\{\beta _{\mathbf{k},\lambda
}\})W_{0}(\{\alpha _{\mathbf{k},\lambda }-\beta _{\mathbf{k},\lambda
}\})d^{2N}\beta .
\]

It is well known that the Wigner function of a state of light may be
obtained by means of the convolution of the (Glauber) P-function and the
Wigner function of the vacuum state when the former exists, that is 
\begin{equation}
W(\{\alpha _{\mathbf{k},\lambda }\})=\int P(\{\beta _{\mathbf{k},\lambda
}\})W_{0}(\{\alpha _{\mathbf{k},\lambda }-\beta _{\mathbf{k},\lambda
}\})d^{2N}\beta .  \label{pw}
\end{equation}
If we identify $W_{1}(\{\alpha _{\mathbf{k},\lambda }\})$ with $P(\{\alpha _{%
\mathbf{k},\lambda }\})$ it is clear that the decomposition of the
stochastic field $\mathbf{E(r,}t)$ into two independent parts requires that
a P-function exists which is positive definite. So, to summarize, the
existence of a positive P-function is a necessary and sufficient condition
for being able to decompose the field into independent signal and ZPF parts.
This is our criterion for a state of light to be classical and it agrees
precisely with the standard quantum-optical definition of classical state.

There is only one classical ''pure'' state, namely the coherent state (the
vacuum state being a particular case). In the ideal situation of a single
mode its P-function is 
\[
P(\alpha )=\delta ^{2}\left( \alpha -a\right) ,
\]
$\delta ^{2}$ being the two-dimensional Dirac's delta. This leads to the
interpretation that the coherent state represents a deterministic
(non-random) signal superimposed to the ZPF. The Wigner function of the
state is given in eq.$\left( \ref{coh}\right) .$

An important kind of classical ''mixed'' state is chaotic light, whose
P-function is 
\begin{equation}
P(\{\alpha _{\mathbf{k},\lambda }\})=\prod_{\mathbf{k},\lambda }\frac{1}{\pi
n_{\mathbf{k},\lambda }}\exp \left( -\frac{|\alpha _{\mathbf{k},\lambda
}|^{2}}{n_{\mathbf{k},\lambda }}\right) ,  \label{cha}
\end{equation}
and the Wigner function 
\begin{equation}
P(\{\alpha _{\mathbf{k},\lambda }\})=\prod_{\mathbf{k},\lambda }\frac{2}{\pi
\left( 2n_{\mathbf{k},\lambda }+1\right) }\exp \left( -\frac{2|\alpha _{%
\mathbf{k},\lambda }|^{2}}{2n_{\mathbf{k},\lambda }+1}\right) .  \label{wcha}
\end{equation}
The quantity $n_{\mathbf{k},\lambda }$ is, in quantum language, the mean
number of photons in the mode $\mathbf{k,}\lambda ,$ as may be easily proved
from the final equality in $\left( \ref{pw0}\right) .$ A particular case of
chaotic is thermal light, where the dependence of $n$ on the frequency $%
\omega =\left| \mathbf{k}c\right| $ is given by Planck\'{}s law.

\section{Entanglement is correlation between quantum fluctuations}

We are in a position to justify the essential conclusion of the present
paper, stated in the title of this section. But I shall not attempt here a
general characterization of entangled states in the Wigner function
formalism, which is a rather formidable problem. It is closely related to
that of determining what density operators correspond to entangled states, a
still open problem actively investigated in the last few years by workers in
quantum information. As said above I shall restrict attention to gaussian
Wigner functions.

\subsection{Gaussian Wigner functions involving two radiation modes}

We are interested in studying a real state of light consisting of radiation
in two separated regions of space (see section 1 for the meaning of real).
Light in every region will contain many modes but, for the sake of clarity,
we shall study an example involving only two modes with amplitudes labelled $%
\alpha _{\mathbf{k},\lambda }$ and $\beta _{\mathbf{k}^{\prime },\lambda
^{\prime }}$, assuming that in all other modes we have just ZPF. A more
physical state would correspond to having two wave packets, the first
(second) containing many modes with wavevectors close to $\mathbf{k}$ $%
\mathbf{(k}^{\prime }\mathbf{)}$\textbf{. }For simplicity we will remove the
subindices and label the amplitudes $\alpha $ and $\beta $ in the following$%
. $

We shall consider a classical two-modes state whose marginal for every mode
corresponds to chaotic light $\left( \ref{cha}\right) $. Its P-function is
gaussian 
\begin{equation}
P_{12}(\alpha ,\beta )=\frac{ab-c^{2}}{\pi ^{2}}\exp \left\{ -a\left| \alpha
\right| ^{2}-b\left| \beta \right| ^{2}+c\left( \alpha \beta ^{*}+\beta
\alpha ^{*}\right) \right\} ,  \label{gausp}
\end{equation}
$a,b$ and $c$ being three real numbers fulfilling $a,b>0,c^{2}<ab$. The
latter condition guarantees that P goes to zero at infinite, a necessary
condition for normalizability. In order to simplify the argument I will
consider a slightly less general state by assuming $a=b$, although the
generalization to $a$ $\neq b$ is trivial. The intensity of the radiation,
above the ZPF, is measured by the integral ( see $\left( \ref{pw0}\right) )$%
\begin{equation}
2n=\int P_{12}(\alpha ,\beta )\left( \left| \alpha \right| ^{2}+\left| \beta
\right| ^{2}\right) d^{2}\alpha d^{2}\beta =\frac{2a}{a^{2}-c^{2}},
\label{n}
\end{equation}
where $n$ represents the mean number of photons per mode.

The Wigner function of this state is given by the convolution with the
vacuum Wigner function ( see $\left( \ref{pw}\right) $.) We shall write it
in terms of $n$ and a correlation parameter, $x,$ defined by 
\begin{equation}
x=\frac{2c}{2a+a^{2}-c^{2}}.  \label{x}
\end{equation}
We get 
\begin{equation}
W_{12}(\alpha ,\beta )=\frac{1-x^{2}}{\pi ^{2}A^{2}}\exp \left\{ -A\left[
\left| \alpha \right| ^{2}+\left| \beta \right| ^{2}-x\left( \alpha \beta
^{*}+\beta \alpha ^{*}\right) \right] \right\} ,A\equiv \frac{2}{\left(
2n+1\right) \left( 1-x^{2}\right) }\smallskip  \label{wnx}
\end{equation}
This Wigner function represents a classical state if $\left| c\right| <a,$
which implies a restriction on the range of values of x. In fact, from $%
\left( \ref{n}\right) $ and $\left( \ref{x}\right) $ we obtain 
\[
a=\frac{4n}{4n^{2}-x^{2}\left( 2n+1\right) ^{2}}.
\]
The condition $a>0$ leads to

\begin{equation}
classical:n\geq 0,\left| x\right| <\frac{2n}{2n+1}.  \label{class}
\end{equation}

A transparent interpretation of these results emerges in our approach.
Remembering eq.$\left( \ref{pw}\right) ,$ we see that $\left( \ref{wnx}%
\right) $ is the probability distribution of the sum of two random variables
representing the signal and the ZPF, respectively. In the signal, whose
distribution is $\left( \ref{gausp}\right) ,$ there is correlation between
the two modes. In contrast in the ZPF, whose distribution is $\left( \ref
{wvac}\right) ,$ all modes are uncorrelated (the distribution is a product
of single-mode terms). In summary, we have \emph{\ a correlation involving
the signal but not the ZPF. }This should be called a \emph{classical
correlation.}

The constraints for the state $\left( \ref{wnx}\right) $ to be real are
weaker than $\left( \ref{class}\right) $, namely 
\[
\mathit{real}:n\geq 0,\left| x\right| <1.
\]
(Another condition for the state to be real is that $W$ may be obtained from
a density operator via $\left( \ref{wro}\right) $, which is true in this
case but will not be proved here). On the other hand the marginal of $\left( 
\ref{wnx}\right) $, 
\[
W_{1}(\alpha )=\frac{2}{\pi \left( 2n+1\right) }\exp \left\{ -\frac{2\left|
\alpha \right| ^{2}}{\left( 2n+1\right) }\right\} ,
\]
represents a classical state. I shall call \emph{entangled} any state of two
modes which is not classical but has classical marginals, that is

\begin{equation}
entangled:n\geq 0,\frac{2n}{2n+1}<\left| x\right| <1.  \label{entangled}
\end{equation}
(I propose this as a sufficient condition, not excluding the possibility of
entangled states with nonclassical marginals). We see that such states
involve a correlation between the two modes which is larger than the
correlation of any classical state (given by $\left( \ref{class}\right) ).$
This is because not only the signal but also \textit{the ZPF or ''quantum
vacuum fluctuations'' are correlated in the entangled state.}

\textit{\ ''}Two-photon entanglement'' similar to this one (although
involving many modes) occurs in the process of parametric down conversion .
In fact, light produced in that way consists of two separated beams whose
Wigner functions are gaussian but the full state is not classical. That is,
every beam alone consists of chaotic light of the type $\left( \ref{wcha}%
\right) $, but the two beams are entangled \cite{pdc1}-\cite{param}.

\textit{\ }The reader may realize that entanglement becomes less relevant
when the intensity $n$ (the mean number of photons) is large because the
interval of $x$ in $\left( \ref{entangled}\right) $ becomes narrow. The
opposite is true if $n$ is small. This is to be expected because large $n$
corresponds to the classical limit.

\subsection{Entanglement and Bell's inequalities}

There is no agreement about the definition of entanglement for mixed quantum
states. A fashionable criterion is the violation of a Bell inequality. In
actual experiments the inequalities tested have never been genuine Bell
inequalities, derived using only general properties of local hidden
variables, but inequalities involving auxiliary assumptions. The violation
of one of such inequalities does not imply the refutation of local realism 
\cite{santos}, a fact qualified as ''existence of loopholes''. Nevertheless
we may consider the experiments as valid tests of entanglement. In practice
any test involves measuring a coincidence detection rate as a function of
some controllable angular parameter, $\phi .$ The inequalities are violated
if the measured coincidence rate, $R_{12}$, is of the form 
\begin{equation}
R_{12}=const.\times \left( 1+V\cos \phi \right) ,  \label{rate}
\end{equation}
with the visibility or contrast, $V,$ greater than some limit, usually 0.71.

The most frequent tests involve polarization correlation. In order to study
polarization we need to take $\mathbf{\alpha }$ and $\mathbf{\beta }$ as
(complex) two-dimensional vectors or, what is equivalent, to double the
number of modes. Introducing the polarization vectors, $\mathbf{e}_{\mathbf{k%
},\lambda },$ see $\left( \ref{field}\right) ,$ the vector amplitude $%
\mathbf{\alpha }$ might be written 
\[
\mathbf{\alpha =}\alpha _{x}\mathbf{e}_{\mathbf{k},1}+\alpha _{y}\mathbf{e}_{%
\mathbf{k},2},\left| \mathbf{\alpha }\right| ^{2}=\left| \alpha _{x}\right|
^{2}+\left| \alpha _{y}\right| ^{2},
\]
and similar for $\mathbf{\beta .}$ The Wigner function will be, instead of $%
\left( \ref{wnx}\right) ,$%
\begin{eqnarray}
W_{12}(\mathbf{\alpha ,\beta }) &=&\frac{\left( 1-x^{2}\right) ^{2}A^{4}}{%
\pi ^{4}}\exp \left\{ -A\left[ \left| \mathbf{\alpha }\right| ^{2}+\left| 
\mathbf{\beta }\right| ^{2}-x\left( \mathbf{\alpha \cdot \beta }^{*}+\mathbf{%
\beta \cdot \alpha }^{*}\right) \right] \right\} ,  \nonumber \\
A &\equiv &\frac{2}{\left( 2n+1\right) \left( 1-x^{2}\right) }.
\label{wvect}
\end{eqnarray}
We shall consider that the amplitudes $\mathbf{\alpha }$ and $\mathbf{\beta }
$ correspond to two polarized light beams arriving at two polarization
analizers at angles $\phi _{1}$ and $\phi _{2}$, respectively. In what
follows we may ignore the ZPF in all modes except those included in $\left( 
\ref{wvect}\right) .$ Consequently we assume that the amplitudes emerging
from the polarizers are given by Malus law, that is 
\[
\lambda =\left( \mathbf{\alpha \cdot u}_{1}\right) ,\mu =\left( \mathbf{%
\beta \cdot u}_{2}\right) ,
\]
the vector $\mathbf{u}_{1}$ having components ($\cos \phi _{1}$, $\sin \phi
_{1}$) and similar for \textbf{u}$_{2}.$ Note that the scalar amplitudes $%
\lambda $ and $\mu $ correspond to modes with polarization in the directions
of $\mathbf{u}_{1}$ and $\mathbf{u}_{2}$, respectively. As said above we
ignore the modes with polarization perpendicular to $\mathbf{u}_{1}$ or $%
\mathbf{u}_{2},$ which contain just ZPF.

The coincidence detection rate in two detectors placed after the polarizers
may be calculated by a straightforward generalization of $\left( \ref{pw0}%
\right) ,$ namely 
\[
R_{12}\propto \int W_{12}\left( \mathbf{\alpha ,\beta }\right) \left( \left|
\lambda \right| ^{2}-\frac{1}{2}\right) \left( \left| \mu \right| ^{2}-\frac{%
1}{2}\right) \text{ }d^{2}\alpha _{x}d^{2}\alpha _{y}d^{2}\beta
_{x}d^{2}\beta _{y}\text{.}
\]
The integration is trivial using eq.$\left( \ref{wvect}\right) $ and we get 
\[
R_{12}\propto n^{2}+\frac{1}{2}\left( n+\frac{1}{2}\right) ^{2}x^{2}\left[
1+\cos \left( 2\phi _{1}-2\phi _{2}\right) \right] .
\]
The visibility is 
\[
V=\frac{\frac{1}{2}\left( n+\frac{1}{2}\right) ^{2}x^{2}}{n^{2}+\frac{1}{2}%
\left( n+\frac{1}{2}\right) ^{2}x^{2}},
\]
which is greater than $\frac{1}{3}$ for entangled states, fulfilling
condition $\left( \ref{entangled}\right) ,$ but it is smaller than $\frac{1}{%
3}$ for classical states, where condition $\left( \ref{class}\right) $ holds
true. In particular, the limit $V=0.71$ may be surpassed if $n$ \TEXTsymbol{<%
} 0.42 ( but this is specific for the gaussian Wigner functions studied in
this paper.)

We see that high visibility is only possible with weak signals. I conjecture
that the signal weakness, combined with the necessity of removing
efficiently the ZPF, gives rise to the difficulties for performing
''loophole-free'' tests of Bell's inequalities. This problem will be studied
elsewhere.

\textbf{Acknowledgement.} I acknowledge financial support from DGICYT,
Project No. PB-98-0191 (Spain).

\end{document}